\documentclass[twocolumn,twocolappendix]{aastex631}

\usepackage{lipsum}

\usepackage{amsmath}
\usepackage{cjhebrew}
\usepackage{lipsum}
\usepackage{float}
\usepackage{afterpage}
\usepackage{makecell}
\usepackage{upgreek}
\usepackage{bm}
\usepackage{ulem}
\usepackage{starfont}
\usepackage[T1]{fontenc}
\usepackage{calligra}
\graphicspath{{./}{figures/}}
\newcommand{\Disco}{{\texttt{Disco}}}

\graphicspath{{./}{figures/}}
\shorttitle{Cooling in Boundary Layers}
\shortauthors{Alexander J. Dittmann}

\begin{document}

\title{The Effects of Cooling on Boundary Layer Accretion}

\correspondingauthor{Alexander J. Dittmann}
\email{dittmann@umd.edu}

\author[0000-0001-6157-6722]{Alexander J.~Dittmann}
\affil{Department of Astronomy and Joint Space-Science Institute, University of Maryland, College Park, MD 20742-2421, USA}

\begin{abstract} 
In many cases accretion proceeds from disks onto planets, stars, white dwarfs, and neutron stars via a boundary layer, a region of intense shear where gas transitions from a near-Keplerian speed to that of the surface. These regions are \textit{not} susceptible to the common magnetorotational and Kelvin-Helmholtz instabilities, and instead global modes generated by supersonic shear instabilities are a leading candidate to govern transport in these regions. This work investigates the dynamics of these systems under a range of thermodynamic conditions, surveying both disk sound speeds and cooling rates. Very fast or very slow cooling has little effect on wave dynamics: in the fast-cooling limit, waves propagate in an effectively isothermal manner, and in the slow limit wave propagation is effectively adiabatic. However, when the cooling timescale is comparable to the wave period, wave damping becomes extreme. In cases with intermediate cooling rates, mass and angular momentum transport can be suppressed by orders of magnitude compared to isothermal and uncooled cases. Cooling in accretion disks leads to a preference for wavenumbers near and below the Mach number of the disk; the corresponding lower frequencies can (in non-isothermal systems) couple to gravity modes within the star, potentially causing low-frequency variability such as dwarf nova and quasi-periodic oscillations in accreting systems.

\end{abstract}
\keywords{Accretion (14); Hydrodynamics (1963); Hydrodynamical simulations (767); Compact binary stars (283) }

\section{Introduction}

Accretion disks form throughout the universe, as compact objects strip material from the envelopes of their companions in binary systems, around supermassive black holes in the centers of galaxies, and around nascent stars and planets. 
These disks are often large reservoirs of angular momentum, some of which they transfer to their central objects, often spinning them up to higher angular velocities. Accretion disks also feed material onto the surfaces of compact stellar remnants, fueling thermonuclear explosions on the surfaces of some accreting white dwarfs and neutron stars. However, the mechanism by which these objects actually accrete --- that is to say, how gas actually traverses between the disk and the surface --- is not well understood.

Boundary layers are an essential component of accretion onto both nascent stars \citep[for example, of the T Tauri and FU Orionis types,][]{1993ApJ...415L.127P} and stellar remnants in cataclysmic variables \citep[e.g.,][]{1978A&A....63..265K,1993Natur.362..820N} and low-mass X-ray binaries \citep[e.g.,][]{1999AstL...25..269I,2003A&A...410..217G}.\footnote{Accretion only proceeds through a boundary layer when the magnetic field strength of the accreting object is insufficient to channel the flow along magnetic field lines \citep[e.g.,][]{1978ApJ...223L..83G}.} These thin interfaces near the stellar surface are thought to produce X-ray and extreme ultraviolet emission in cataclysmic variables, for example \citep[e.g.,][]{1997ApJ...475..812M,1985ApJ...292..550P}. These accreting systems can exhibit variability on non-orbital timescales, including quasi-periodic oscillations and dwarf nova oscillations \citep[e.g.,][]{2005ASPC..330..227W}, which may be related to instabilities or waves in the accretion disk \citep[e.g.,][]{2017ApJ...835..238B}. 

Despite the ubiquity of accretion disks, the physical mechanisms for accretion onto the surfaces of these objects are understood to only a nebulous degree. One problem is that, save for black holes, most accreting objects have surfaces, which necessarily rotate at sub-Keplerian speeds. Thus, the angular velocity of the gas must increase with radius, rendering the fluid \textit{stable} to the magnetorotational instability \citep{velikhov59,1960PNAS...46..253C} thought to stimulate transport in many accretion disks \citep{1998RvMP...70....1B}. Many boundary layer models have circumvented the issue of a physical transport mechanism through the assumption of an ad hoc Navier-Stokes viscosity \citep[e.g.][]{1995ApJ...442..337P,1999AstL...25..269I,2004ApJ...616L.155P,2009ApJ...702.1536B,2021ApJ...921...54D}. Another challenge is that accretion disks tend to flow at highly supersonic velocities near the stellar surface, rendering them stable against the Kelvin-Helmholtz instability in the direction of the flow \citep{1958JFM.....4..538M}, limiting the available hydrodynamic mechanisms for driving turbulence. 

A path forward was charted by \citet{2012ApJ...760...22B}, which demonstrated that supersonic shear instabilities \citep{1988MNRAS.231..795G,2012ApJ...752..115B} could generate waves capable of transporting angular momentum through the boundary layer. Since then, a number of works have investigated this mechanism, exploring factors such as magnetic fields, the disk sound speed, and the influence of stellar rotation \citep{2012ApJ...760...22B,2013ApJ...770...67B,2013ApJ...770...68B,2018MNRAS.479.1528B,2021MNRAS.508.1842D,2022MNRAS.509..440C,2022MNRAS.512.2945C}. Although the aforementioned studies assumed isothermal equations of state, \citet{2015A&A...579A..54H} found that sonic instabilities still operated in a simulation using a diffusion approximation of radiative transport, and \citet{2017ApJ...835..238B} showed that acoustic waves in non-isothermal systems can excite incompressible waves in the central object. 

\citet{2016ApJ...817...62P} showed that the same acoustic instabilities operate and transport angular momentum in spreading layers, where material spreads vertically from low to high latitudes after joining the stellar surface. Additionally, \citet{2016ApJ...817...62P} explored disks with a range of cooling rates, finding that the transport in the spreading layer was greatly reduced in cases with intermediate cooling rates, whereas uncooled spreading layers evolved similarly to isothermal ones. The effects of cooling on accretion disks have also been studied in the context of planetary migration: numerous studies have found, both numerically and semi-analytically, that cooling can dramatically affect the propagation of waves in accretion disks driven by embedded planets \citep[e.g.,][]{2020ApJ...892...65M,2020MNRAS.493.2287Z,2024ApJ...961...86Z,2024MNRAS.528.6130Z} when the cooling timescale is of the same order of magnitude as the dynamical timescale of the accretion disk. 

The goal of this paper is to develop a systematic understanding of how cooling affects waves in boundary layers and the resulting transport between disk and central object. Section \ref{sec:methods} introduces the equations relevant to this study, develops some analytical estimates to build intuition about cooled disks, and discusses the numerical methods used to approximate solutions to the equations of hydrodynamics. Section \ref{sec:results} presents the results of a suite of hydrodynamical simulations, surveying a variety of disk sound speeds and cooling rates. Caveats and astrophysical implications are discussed in Section \ref{sec:discussion}, and the work is summarized in Section \ref{sec:summary}.

%\pagebreak
\section{Fluid Dynamics}\label{sec:methods}
The present investigation is limited to two-dimensional studies of accretion disks, and the consequences of this choice are discussed in Section \ref{sec:caveatEmptor}.
The equations governing two-dimensional (vertically-integrated) inviscid hydrodynamics are those of mass and momentum conservation, 
 \begin{align}
\partial_t\Sigma + \nabla\cdot(\Sigma \mathbf{v})=0, \label{eq:continuity}\\
\partial_t(\Sigma \mathbf{v}) + \nabla\cdot(\Sigma\mathbf{v}\mathbf{v}+P\mathbb{I})=-\Sigma\nabla\Phi,\label{eq:momentum}
\end{align}
where $\mathbf{v}$ is the fluid velocity, $\mathbb{I}$ is the identity tensor, and $\Phi$ is a gravitational potential. These equations must be supplemented with an additional equation to relate the surface density $\Sigma$ to the vertically-integrated scalar pressure $P$. The simplest of these relations is an isothermal equation of state, so that
\begin{equation}
P = c_s^2\Sigma, 
\end{equation}
where $c_s$ is the \emph{isothermal} sound speed. One may also assume an ideal gas equation of state
\begin{equation}\label{eq:lociso}
P=(\gamma-1)e\Sigma,
\end{equation}
where $\gamma$ is the adiabatic index and $e$ is the specific internal energy. If we consider adiabatic variations, such that the fluid entropy $ S\propto\log{P/\Sigma^\gamma}$ is conserved in a Lagrangian sense ($dS/dt=0$), then the system of fluid equations can instead be closed according to
\begin{equation}\label{eq:adiabatic}
\partial_te-\frac{P}{\Sigma^2}\partial_t\Sigma=0.
\end{equation}

However, reality is rarely precisely isothermal or adiabatic, and many fluids experience heating and cooling. In order to probe the effects of these processes on boundary layer dynamics, I have employed a simple Newtonian cooling source term, which relaxes the fluid towards some fixed temperature profile ($e_0$) on a characteristic ``cooling'' timescale ($t_c$), 
\begin{equation}\label{eq:cooling}
\partial_te-(\gamma-1)\frac{e}{\Sigma}\partial_t\Sigma=-\frac{e-e_0}{t_c}.
\end{equation}
Notably, when $t_c\ll(\partial_t\Sigma/\Sigma)^{-1},$ the specific energy profile of the fluid will be virtually identical to $e_0$ and the system will be effectively isothermal. Conversely, when $t_c\gg(\partial_t\Sigma/\Sigma)^{-1},$ cooling will have a negligible effect and the dynamics will proceed nearly adiabatically. 
To build further intuition for these processes, we will examine a very simplified system in Section \ref{sec:cartdisp}, and then linearly perturbed accretion disks in Section \ref{sec:diskdisp}. Section \ref{sec:numerics} details the numerical methodology applied to these systems in Section \ref{sec:results}.  

\subsection{Waves in One Dimension}\label{sec:cartdisp}
For the sake of building intuition, it is useful to first consider simple waves. The linearized equations governing an adiabatically evolving fluid in a one-dimensional Cartesian domain can be written in terms of the fluid density ($\rho$) and pressure ($p$) as
\begin{align}
\partial_t\delta\rho+\rho_0\partial_x\delta v=0,\\
\partial_t\delta v+ \frac{1}{\rho_0}\partial_x\delta p =0,\\
\gamma c_s^2\partial_t\delta\rho - \partial_t\delta p =0, 
\end{align}
assuming a static and uniform background such that the total fluid quantities can be decomposed into constant components and fluctuating perturbations according to $\rho=\rho_0+\delta\rho$, $p=p_0+\delta p$, and $v=\delta v$. Assuming simple perturbations of the form $\delta\sim{\rm exp}(ikx-i\omega t)$, one can recover the dispersion relation for waves in a uniform static fluid,
\begin{equation}\label{eq:basicdisp}
\omega(\omega^2-\gamma c_s^2k^2)=0,
\end{equation}
where $\omega=0$ is the familiar entropy wave and $\omega^2=\gamma c_s^2k^2$ describes linear sound waves. 

If we instead consider cooling as described by Equation (\ref{eq:cooling}), then the linearized energy equation becomes 
\begin{equation}
c_s^2(i\omega\gamma - t_c^{-1})\delta\rho - (i\omega - t_c^{-1})\delta p = 0,
\end{equation}
and the resulting dispersion relation is
\begin{equation}\label{eq:coolsounddisp}
\omega^2-c_s^2k^2 - i\omega t_c(\omega^2-\gamma c_s^2k^2)=0.
\end{equation}
In the limit of instantaneous cooling ($\omega t_c\rightarrow0$), Equation (\ref{eq:coolsounddisp}) describes linear isothermal sound waves, and in the limit of no cooling ($\omega t_c\rightarrow\infty$) Equation (\ref{eq:coolsounddisp}) reduces to Equation (\ref{eq:basicdisp}). However, at intermediate cooling timescales the evolution is more complex. It is straightforward to interpret how the dimensionless quantity $\omega t_c$ governs wave dynamics: If cooling occurs very rapidly compared to a single oscillation, the fluid is effectively isothermal; on the contrary, when cooling timescales are much longer than the oscillatory timescales, the fluid is essentially adiabatic.

\begin{figure}
\includegraphics[width=\linewidth]{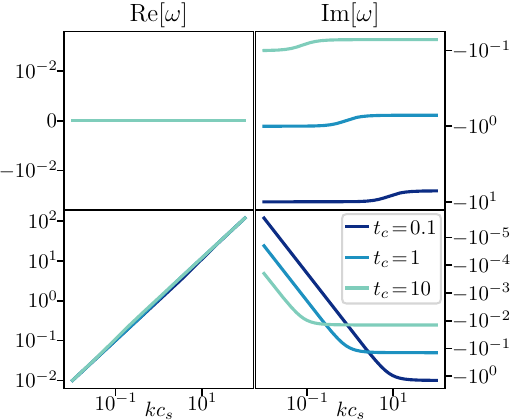}
\caption{The real and imaginary parts of the wave frequency for the modes supported by Equation (\ref{eq:coolsounddisp}). The entropy wave is shown in the top row and the sound wave is shown in the second row. For simplicity, values are only shown for $k>0$, although ${\rm Im}[\omega]$ is symmetric about $k=0,$ depending only on $|k|$. Similarly, only one of the sound waves is plotted, the other having the same imaginary part but opposite real parts.}
\label{fig:coolsound}
\end{figure}

Although cubic polynomials such as Equation (\ref{eq:coolsounddisp}) are analytically tractable, the exact expression for $\omega(k)$ is in this case too long to provide insight. Instead, Figure \ref{fig:coolsound} displays the real and imaginary parts of each wave mode for three choices of the cooling timescale. Notably, ${\rm Im}[\omega]\leq0$, confirming our intuition that cooling should damp perturbations. Concerning the entropy mode, ${\rm Im}[\omega]\rightarrow 0$ as $t_c\rightarrow\infty$ and ${\rm Im}[\omega]\rightarrow -\infty$ as $t_c\rightarrow 0$, capturing the correct behavior in both the isothermal and adiabatic limits. The imaginary parts of the frequencies of sound waves tend towards zero superlinearly for $kc_s\lesssim t_c^{-1}$, as the flow is effectively isothermal at low frequencies and wavenumbers; at higher frequencies, ${\rm Im}[\omega]$ becomes approximately constant, and ${\rm Im}[\omega]/{\rm Re}[\omega]\rightarrow 0$. For any finite wavenumber, the isothermal limit is recovered as $t_c\rightarrow0$, as the break wavenumber (that above which the imaginary part asymptotes) tends towards infinity and the pre-break imaginary part of the dispersion relation tends towards zero; and the adiabatic limit is recovered as $t_c\rightarrow\infty$, as the break wavenumber and post-break amplitude tend towards zero.

\subsection{Waves in Disks}\label{sec:diskdisp}
For this study of disks, I will take the background surface density to be constant and the angular velocity profile to be an arbitrary function of radius $\Omega(r)$. 
In this case, the linearized fluid equations, including cooling, can be written as 
\begin{align}
\partial_t\frac{\delta\Sigma}{\Sigma} + \Omega\partial_\phi\left(\frac{\delta\Sigma}{\Sigma}\right) +\partial_r\delta v_r + \frac{\delta v_r}{r}+\frac{1}{r}\partial_\phi\delta v_\phi\!=\! 0\\
\partial_t\delta v_r + \Omega\partial_\phi\delta v_r - 2\Omega\delta v_\phi + \partial_r\left(\frac{\delta P}{\Sigma}\right) \!=\!0\\
\partial_t\delta v_\phi +\Omega\partial_\phi\delta v_\phi \!+\!\partial_r(\!r\Omega\!)\delta v_r \!+\! \Omega\delta v_r \!+\! \frac{1}{r}\partial_\phi\!\left(\!\!\frac{\delta P}{\Sigma}\!\right)\! \!=\!0\\
c_s^2\!\left(\!\gamma(\partial_t\!+\!\Omega\partial_\phi)\!+\frac{1}{t_c}\right)\!\frac{\delta\Sigma}{\Sigma}\!-\!\left(\frac{1}{t_c}\!+\!\partial_t\!+\!\Omega\partial_\phi\!\right)\!\!\frac{\delta P}{\Sigma}\!=\!0.
\end{align}
After defining the squared radial epicyclic frequency and $\kappa^2\equiv2\Omega(2\Omega+r\partial_r\Omega)$,taking perturbations of the form \mbox{$\delta\sim{\rm exp}[i \int^r r'k_r(r') + im(\phi - \Omega_pt)$]} for a perturbation with some azimuthal pattern speed $\Omega_p$, and defining the redshifted frequency $\tilde{\omega}\equiv m(\Omega_p-\Omega)$, the resulting dispersion relation is

\begin{equation}\label{eq:cooldiskdisp}
\begin{split}
\left[\tilde{\omega}^2-\kappa^2-c_s^2k_r^2-\frac{c_s^2m}{r^2}\left(m+2\frac{\Omega}{\tilde{\omega}}\right)\right]\\
-i\tilde{\omega}t_c\left[\tilde{\omega}^2-\kappa^2-\gamma c_s^2k_r^2-\frac{\gamma c_s^2m}{r^2}\left(m+2\frac{\Omega}{\tilde{\omega}}\right)\right]\\
+\frac{c_s^2k_rm}{r}\left(\frac{1}{m}+\frac{\kappa^2}{2\Omega\tilde{\omega}}-\frac{2\Omega}{\tilde{\omega}}\right)\left(i+\gamma\tilde{\omega}t_c\right)=0.
\end{split}
\end{equation}
Taking the limit of $t_c\rightarrow0$, Equation (\ref{eq:cooldiskdisp}) reduces to 
\begin{equation} \label{eq:isodiskdisp}
\begin{split}
\tilde{\omega}^2-\kappa^2-c_s^2k_r^2-\frac{c_s^2m}{r^2}\left(m+2\frac{\Omega}{\tilde{\omega}}\right)\\
+i\frac{c_s^2k_rm}{r}\left(\frac{1}{m}+\frac{\kappa^2}{2\Omega\tilde{\omega}}-\frac{2\Omega}{\tilde{\omega}}\right)=0,
\end{split}
\end{equation}
which matches Equation (A16) of \citet{2013ApJ...770...67B}.

Qualitatively, Equation (\ref{eq:cooldiskdisp}) bears many similarities to Equation (\ref{eq:basicdisp}); by taking the $t_c\rightarrow\infty$ limit, one recovers nearly the same dispersion relation as the $t_c\rightarrow0$ limit, but with the addition of an $\tilde{\omega}=0$ mode and and the replacement of the isothermal sound speed by the adiabatic sound speed.

\begin{figure}
\includegraphics[width=\linewidth]{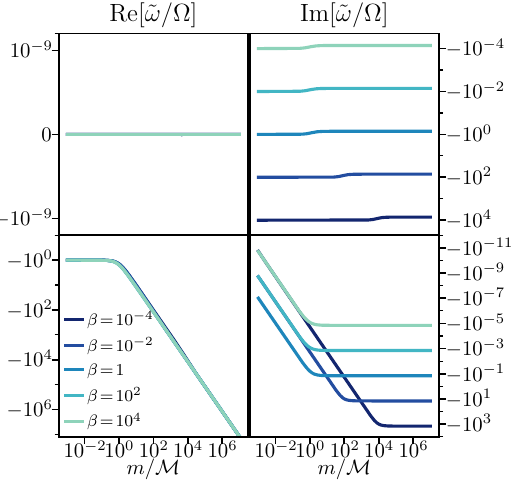}
\caption{The real and imaginary parts of the solutions to Equation (\ref{eq:simplecooldisk}). For simplicity, this plot only includes the solution branch with $\Omega_p<\Omega,$ and hence ${\rm Re}[\tilde{\omega}]<0$. Qualitatively, the dispersion relations are very similar to those of sound waves, but with a low-$m$ cutoff, below which the damping rate due to cooling precipitously declines. Qualitatively, this should cause higher-$m$ modes to die out over time unless continuously excited. }
\label{fig:diskdisp}
\end{figure}

Taking the limits $m\ll r\Omega/c_s$ and $k_r\gg m/r$, Equation (\ref{eq:isodiskdisp}) reduces to the standard ``tight-winding'' dispersion relation for spiral density waves \citep[e.g.][]{1978ApJ...222..850G}
\begin{equation}
\tilde{\omega}^2 - \kappa^2 - k_r^2c_s^2 = 0.
\end{equation}
As described in \citet{2013ApJ...770...67B}, the tight-winding approximation is rarely applicable to wave transport in boundary layers and instead a high-$m$ ``loose-winding'' approximation is more suitable, assuming $m\gg1$ and $m\gg rk_r$. Under these assumptions Equation (\ref{eq:isodiskdisp}) becomes far more tractable, yielding
\begin{equation}
\tilde{\omega}^2=c_s^2\frac{m^2}{r^2} + \kappa^2.
\end{equation}
It is worth noting that in many cases one can combine the ``tight-winding'' and ``loose-winding'' regimes by instead taking $m^2/r^2\rightarrow m^2/r^2 + k_r$ \citep[e.g.,][]{2013ApJ...770...67B}, but here I will retain only $m$ for the sake of brevity.  
In the same spirit, Equation (\ref{eq:cooldiskdisp}) reduces in this approximation to 
\begin{equation}\label{eq:simplecooldisk}
\tilde{\omega}^2 \!-\kappa^2 \!- c_s^2\frac{m^2}{r^2}-i\tilde{\omega}t_c(\tilde{\omega}^2-\kappa^2-\gamma c_s^2\frac{m^2}{r^2})\!=0,
\end{equation}
which can be non-dimensionalized, also defining the Mach number as $\mathcal{M}=r\Omega/c_s$ and parameterizing the cooling timescale as $t_c=\beta/\Omega$, as 
\begin{equation}\label{eq:simplecooldiskND}
\frac{\tilde{\omega}^2}{\Omega^2} \!-\frac{\kappa^2}{\Omega^2} \!- \frac{m^2}{\mathcal{M}^2}-i\beta\frac{\tilde{\omega}}{\Omega}(\frac{\tilde{\omega}^2}{\Omega^2}-\frac{\kappa^2}{\Omega^2}-\gamma\frac{m^2}{\mathcal{M}^2})\!=0.
\end{equation}
This is qualitatively similar to Equation \ref{eq:coolsounddisp}, but with a low-frequency cutoff such that $\tilde{\omega}\rightarrow\kappa$ when $m\ll \mathcal{M}$.  Some solutions to Equation (\ref{eq:simplecooldiskND}) --- fixing $\gamma=1.4$, $\kappa/\Omega = 1,$ and holding $\beta$ constant --- are shown in Figure \ref{fig:diskdisp}. Pronounced wave damping is again caused by the cooling term, particularly when $\tilde{\omega}\sim t_c^{-1}$.

In the high-wavenumber limit ($m/\mathcal{M}\gg \tilde{\omega}/\Omega$), longer cooling timescales universally lead to less damping; in this limit damping is again appreciable in an absolute sense, but over the course of one wave period damping becomes negligible. At lower wavenumbers, $m/\mathcal{M}\lesssim \kappa/\Omega$, disks with $\beta=\Omega/\kappa$ experience the most prominent damping. For any finite cooling timescale, it is then natural that thinner, higher Mach number disks should support higher wavenumber modes than thicker disks, and that over time higher wavenumber modes should die out in favor of lower wavenumber modes. Additionally, high-wavenumber modes should experience less damping in disks with longer cooling timescales. 

\subsection{Cooling in Astrophysical Disks}
In astrophysical systems, the cooling timescale can take a wide range of values, from many times shorter than the dynamical timescale to many times longer. Taking optically thick accretion disks as an example, the emergent flux from some patch of the disk can be approximated as $\sigma T_{\rm eff}^4\approx\sigma T^4/\tau$, where $\sigma$ is the Stefan-Boltzmann constant and $\tau$ is the optical depth. In a radiation-dominated disk, such as one belonging to a neutron star, the energy density is $4\sigma T^4/c$, and thus the ratio of the cooling timescale to the dynamical timescale is $\beta\sim \tau c_s/c$. In the inner regions of disks around compact objects, $\tau$ can be quite large and $c_s/c$ may be on the order of a few percent, leading to $\beta\gg1$. In the outer, less relativistic portions of the disk, lower sound speeds lead to $\beta\ll1,$ although generally $\beta$ will also be proportional to the ratio of the total energy density to the radiation energy density. Protoplanetary disks are also expected to have a wide range of cooling timescales, from $\beta\sim10^{5}$ in their inner regions to $\beta\sim10^{-2}$ in their outer regions \citep[e.g.,][]{2015ApJ...813...88Z}.

Although the simple Newtonian cooling scheme adopted here is suited to gauging the general effects of cooling in a variety of disks, it cannot emulate some potentially important features of astrophysical disks. For example, in optically thick disks radiative diffusion leads to a dependence of the cooling timescale on the azimuthal wavenumber \citep[e.g.][]{2015ApJ...811...17L,2020ApJ...904..121M}. External irradiation often strongly affects protoplanetary disks \citep[e.g.][]{2015A&A...575A..28B}, and the thermal structure of accretion disks around compact objects can be governed by the coupling between ions and electrons by Coulomb collisions \citep[e.g.,][]{2024MNRAS.527.2895B}. Nevertheless, the present investigation will serve to shed light on the general effects of cooling, particularly on mass and angular momentum transport. 

\subsection{Numerical Methods}\label{sec:numerics}
To study the problem of transport in astrophysical boundary layers in more detail, I have conducted a suite of hydrodynamical simulations using the moving-mesh finite-volume code \Disco{} \citep{2016ApJS..226....2D}.\footnote{Specifically, I used the version \url{https://github.com/NYU-CAL/Disco/tree/ryan}, which includes additional optimizations and in-situ diagnostics.} In addition to Equations (\ref{eq:continuity}) and (\ref{eq:momentum}), \Disco{} solved either Equation (\ref{eq:lociso}) or a modified version of the energy equation 
\begin{equation}
\begin{split}
\partial_t\!\!\left(\frac{1}{2}\Sigma\tilde{\mathbf{v}}^2\!+\!e\!\right)\!+\!\nabla\!\!\cdot\!\!\left(\!\!\!\left(\frac{1}{2}\Sigma\tilde{\mathbf{v}}^2+e+P\right)\!\!\mathbf{v}\!\!\right)+\Sigma \mathbf{v}\cdot\nabla\Phi\\=r\Sigma v_r\left(\Omega_K^2(r)-r\tilde{\Omega}\partial_r\Omega_K(r)\right)-\frac{\Sigma\Omega_K}{\beta}(e-e_0),
\end{split}
\end{equation}
where $\tilde{\mathbf{v}}=\{v_r, r(\Omega-\Omega_K)\}$ and cooling is implemented along the lines of Equation (\ref{eq:cooling}), setting $t_c=\beta/\Omega_K$. In this work, $\Phi=-GM_*/r$ is the gravitational potential due to the central object of mass $M_*$ and $\Omega_K=\sqrt{GM_*/r^3}$ is the Keplerian angular frequency. This particular expression of the energy equation subtracts off a static Keplerian velocity field which can reduce the effects of round-off errors in the accretion disk, although for the disks studied in this work this choice is not consequential. Each non-isothermal simulation used an adiabatic index of $\gamma=7/5$. 

Each simulation was initialized with an angular velocity profile
\begin{equation}
\Omega(r)\!=\!
\left\{
\begin{array}{ll}
      r^{-3/2}    &\! r > r_*+\delta r\\
      \!\!\!\left[\!\left(r_*+\delta r\right)^{-3/2}\right]\!\frac{r+\delta_r-r_*}{2\delta r}\! & \!r_*\!-\!\delta r \leq \!r\! \leq \!r\! + \!\delta r\\
      0 &\! r < r_*-\delta r, \\
\end{array}
\right.
\end{equation}
an isothermal temperature profile, and a surface density profile such that the system began in hydrostatic equilibrium, satisfying
\begin{equation}
\frac{1}{\Sigma}\frac{P}{dr}=-\frac{d\Phi}{dr}+\Omega^2r.
\end{equation}
In these equations, $r_*$ is the radial location of the stellar surface, $\delta r=r_*/100$ is the initial half-width of the shear layer. The sound speed of each simulation defines a characteristic Mach number at the stellar surface, $M_*\equiv r\Omega_0/c_s,$ where $\Omega_0$ is the Keplerian angular velocity at the stellar surface and I investigated disks with $\mathcal{M}_*=\{6,8,10\}$. These disk sound speeds may be fairly appropriate to circumplanetary disks \citep[e.g.][]{2007astro.ph..1485A,2021ApJ...921...54D}, although the accretion disks around accretion white dwarfs and neutron stars are thought to typically be much cooler, by factors of $\gtrsim10$.\footnote{For example, a typical accreting white dwarf might have a temperature of $T\sim10^5$ K, a radius of $r_*\sim10^9$, and a mass of $M\sim0.6M_\odot$, suggesting $\mathcal{M}\sim100$. } It is not feasible to simulate such cold disks at present, but some studies have suggested that $\mathcal{M}_*\sim10$ might be sufficient to capture a qualitative picture of boundary layer dynamics in thin disks \citep[e.g.][]{2012ApJ...760...22B,2022MNRAS.509..440C,2022MNRAS.512.2945C}. 

\begin{figure*}
\includegraphics[width=\linewidth]{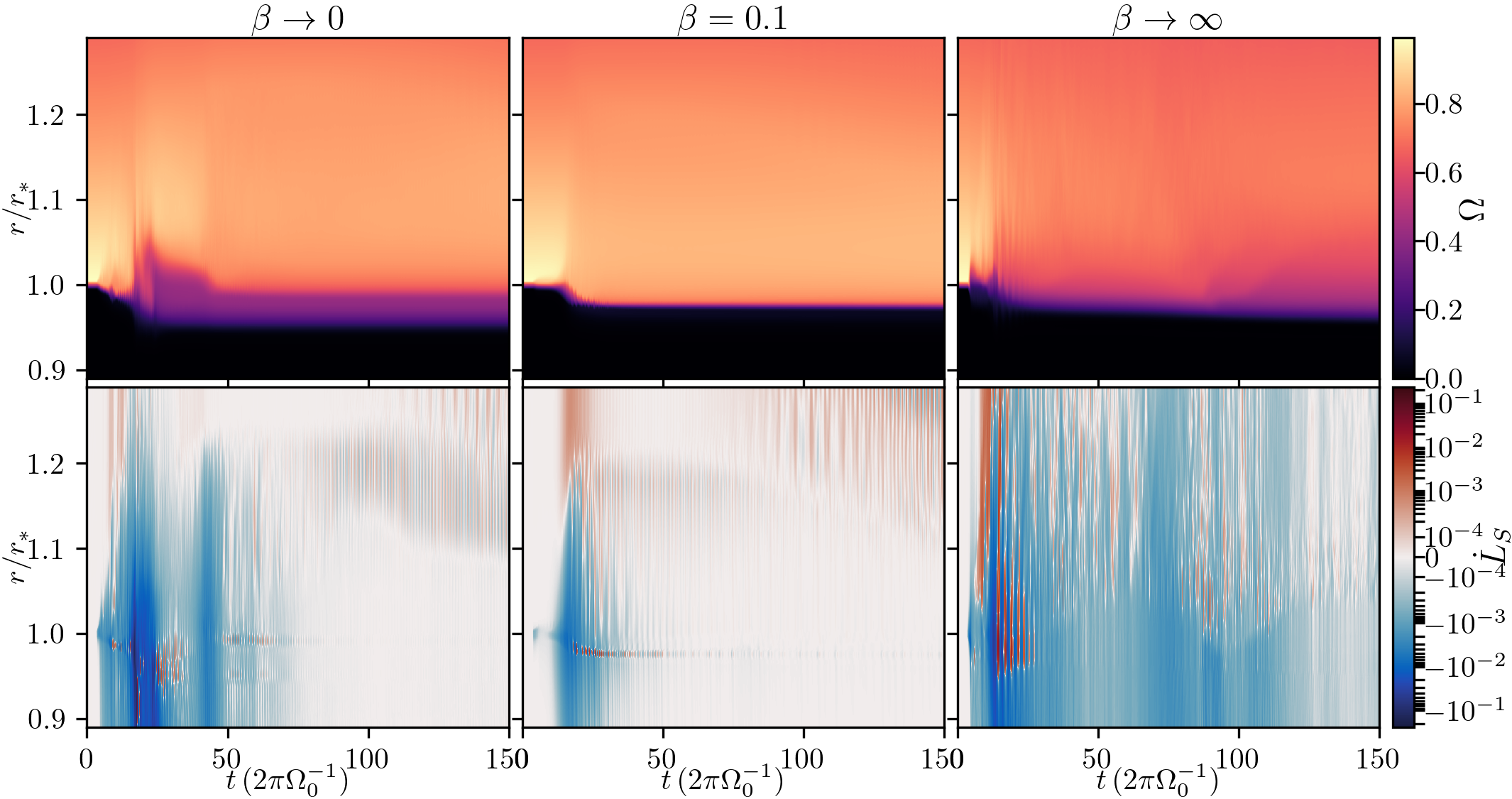}
\caption{The top row plots the azimuthally-averaged (mass-weighted) angular velocity as a function of radius and time, and the bottom row plots the angular momentum current through the star-disk system due to stresses. The first, second, and third columns plot results for isothermal, $\beta=0.1,$ and $\beta\rightarrow\infty$ disks respectively, in each case setting $\mathcal{M}_*=8$. The simulations used to construct this figure, and only this figure, employed a reduced azimuthal range of $\phi\in[0,\pi/4]$ to save disk space when saving simulation outputs at a high cadence.}
\label{fig:spacetime}
\end{figure*}

\begin{figure*}
\includegraphics[width=\linewidth]{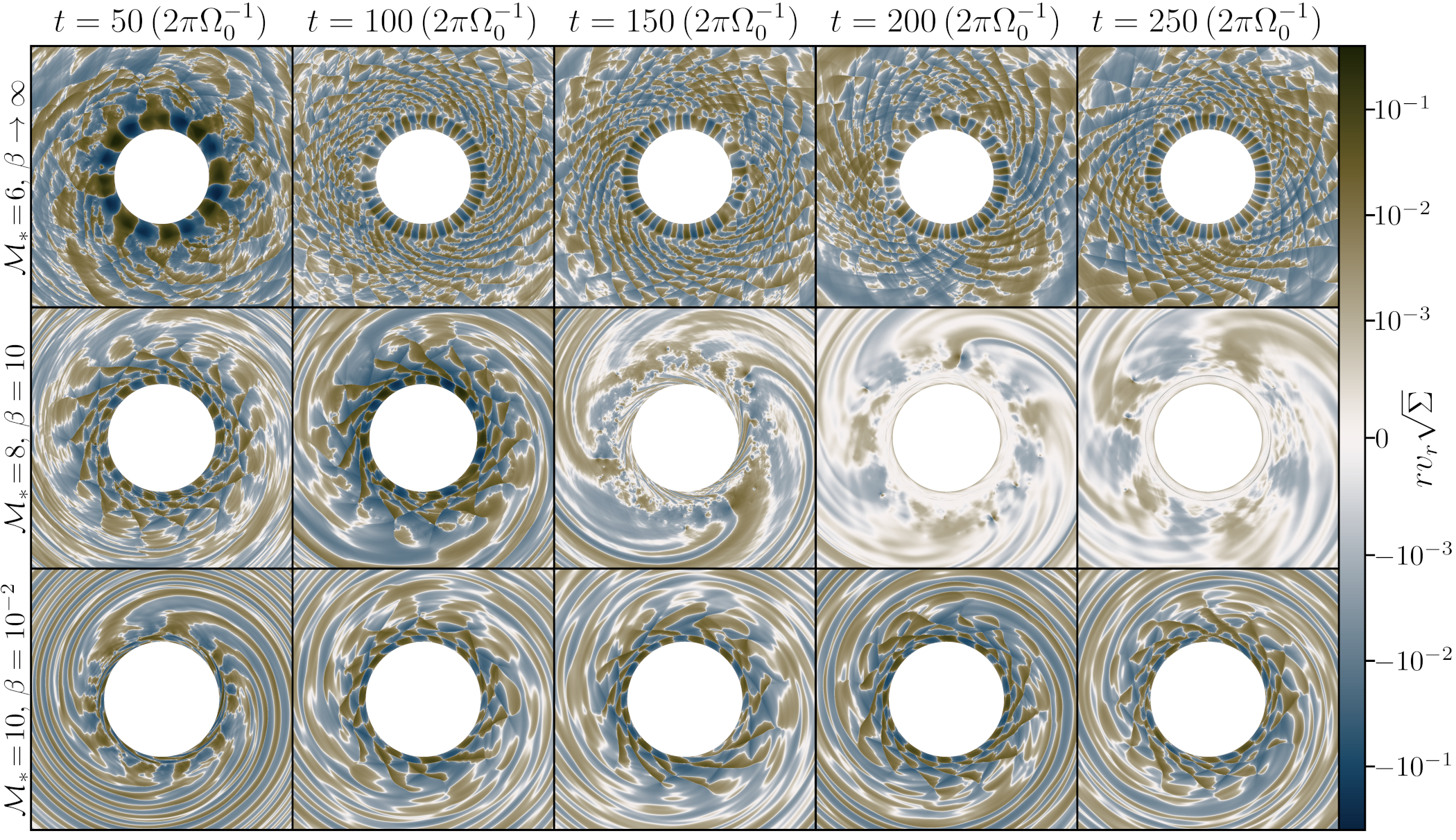}
\caption{Snapshots of the wave quasi-action $rv_r\sqrt{\Sigma}$ over time for disks with varying Mach numbers and cooling rates. Each panel is $4r_*$ tall and wide. Waves in disks with nearer-to-unity cooling rates die out more quickly, accompanied by a preference for lower-$m$ features. }
\label{fig:timesurvey}
\end{figure*}

\begin{figure*}
\centering
\includegraphics[width=0.99\linewidth]{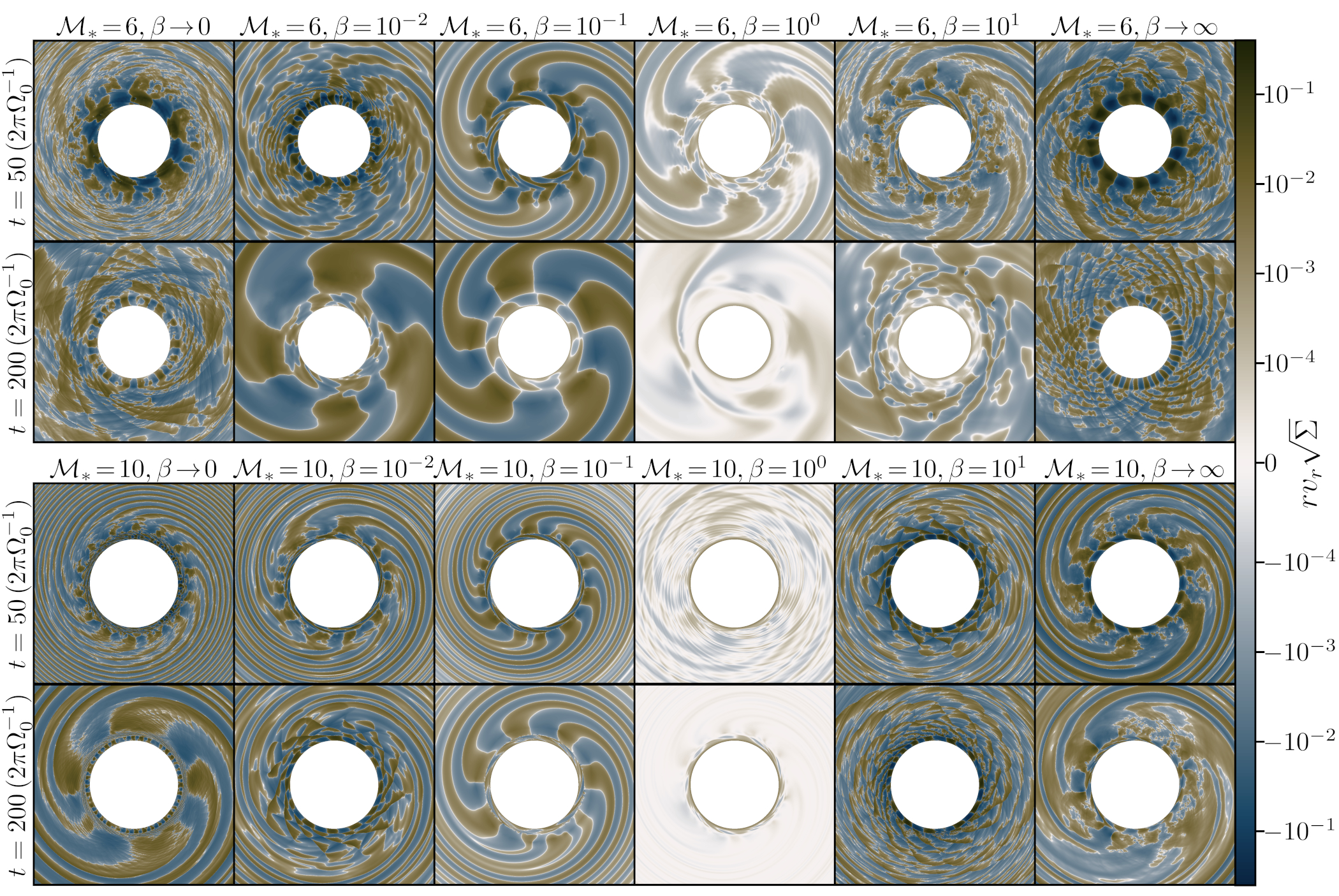}
\caption{Snapshots of the wave quasi-action ($rv_r\sqrt{\Sigma}$) over time for disks with $\mathcal{M}_*=6$ and $\mathcal{M}_*=10$, illustrating mode morphology at early and late times in each case. Over time modes tend towards lower wavenumbers, and cooling rates closer to $\sim1$ lead to the greatest wave suppression.
Each panel is $4r_*$ tall and wide.}
\label{fig:machsurvey}
\end{figure*}

\begin{figure*}
\centering
\includegraphics[width=\linewidth]{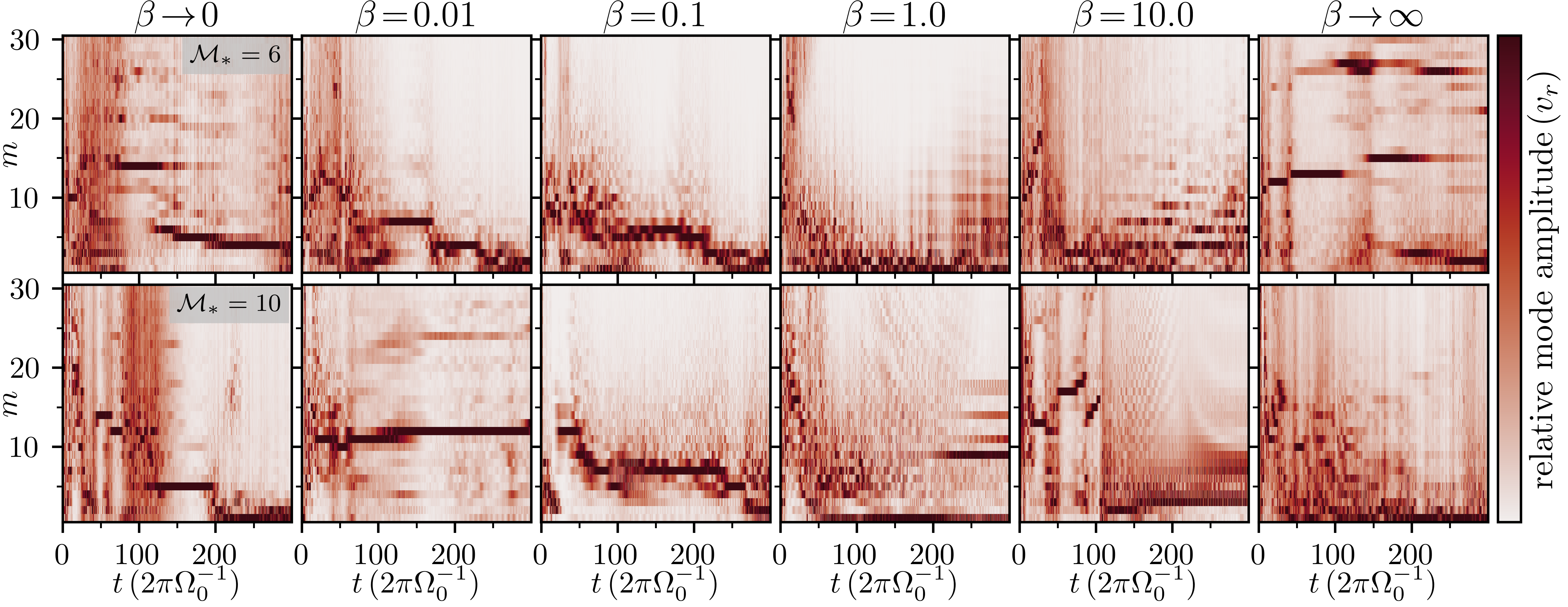}
\caption{The amplitude of modes over time with azimuthal wavenumbers between $m=1$ and $m=30$ over time in $\mathcal{M}_*=6$ (top row) and $\mathcal{M}_*=10$ (bottom row) under a variety of thermodynamic treatments. At each point in time the mode amplitudes have been normalized so that the maximum amplitude is constant over time, illustrating the morphological prominence of each mode rather than its strength in physical terms. These amplitudes were calculated from the Fourier decomposition of the radial velocity in the disk, averaged over the range $1.3 \leq r/r_* \leq 1.5$. Cooling tends to suppress modes with wavenumbers $m\gtrsim \mathcal{M}_*$.}
\label{fig:diskmodes}
\end{figure*}

The outer boundary of each simulation was fixed at $3r_*$, and the inner boundary was adjusted based on $\mathcal{M}_*$ to achieve a density contrast of $\sim10^6$ between the surface density at the inner boundary and the surface density of the disk. The radial cells numbered $N_r=2048$ and were logarithmically spaced, and number of azimuthal cells was adjusted to keep the cell aspect ratio as close to unity as possible. This led to inner boundaries located at $r_{\rm in}/r_*=\{0.7235,0.8243,0.8799\}$ and azimuthal resolutions of $N_\phi=\{9050,9956,10494\}$ for $\mathcal{M}_*=\{6,8,10\}$. For each disk temperature, I carried out an isothermal simulation (effectively $\beta\rightarrow0$, in this case also taking $\gamma=1.0$);\footnote{I have only considered globally isothermal disks in this work. However, if $c_s$ varied spatially, the $\beta\rightarrow0$ limit would lead to a \textit{locally isothermal} disk, with a varying but fixed temperature profile.}  one adiabatic simulation, effectively taking $\beta\rightarrow\infty$; and a set of simulations with finite cooling timescales ($\beta=\{1/100,1/10,1,10\}$), focusing on the regime where cooling is likely to have the most pronounced effects.

Two quantities of interest are the accretion rate through the disk ($\dot{M}$) and the angular momentum current through the disk due to stresses (predominantly waves, $\dot{L}_s$). These are defined as $\dot{M}=2\pi r\bar{\Sigma}\langle v_r\rangle$ and $\dot{L}_S=2\pi r^2\bar{\Sigma}(\langle v_r v_\phi\rangle - \langle v_r\rangle\langle v_\phi\rangle)$, where $\langle ...\rangle$ represents a mass-weighted azimuthal average, and $\bar{\Sigma}$ is the azimuthally averaged surface density. In practice, these are also averaged in time each time step. Additionally, every 100th of an orbital period at the stellar surface ($2\pi\Omega_0^{-1},$ a natural time unit for these systems), I calculated the Fourier coefficients of the radial velocity at each radius to quantify the presence of each mode over time. For purposes of visualizing the morphology of these systems, it will also be useful to examine $rv_r\sqrt{\Sigma}$, which is a proxy for wave action allowing for visualization of modes both within the star and disk. 

\section{Numerical Results} \label{sec:results}
Because the goal of this study is to understand the effects of cooling on boundary layer dynamics, rather than the nature of sonic instabilities themselves, I will focus on transport and long-lived wave modes rather than dwelling on the initial development of instabilities in the boundary layer. Qualitatively, regions of strong supersonic shear between the disk and star\footnote{For the sake of simplicity, I will use ``star'' as a generic stand-in throughout this section for any central accreting object, from protoplanets to neutron stars.} can lead to corotation resonances which can amplify waves upon reflection and partial tunneling; the trapping of these waves leads to instability.\footnote{Specifically, these waves have a conserved action that changes sign at corotation, any partial tunneling leads to amplification of the reflected wave. Waves trapped between two resonances, or between one resonance and a wall, are then unstable. See \citet{2012ApJ...752..115B} and \citet{2012ApJ...760...22B} for a more rigorous analysis of this instability.} 
The two most common modes generated in these systems are often referred to as ``lower'' modes, which correspond to the usual $p$ and $g$ modes of stratified atmospheres \citep[e.g.,][Chapter 2]{2006aofd.book.....V}, and ``upper'' modes, which have the character of the usual waves in disks \citep[e.g.,][]{1978ApJ...222..850G}; a more thorough description is provided in Section 4 of \citet{2013ApJ...770...67B}. \citet{2022MNRAS.509..440C} also identified the development of vortex-driven modes, where a vortex in the boundary layer sources a spiral wave the propagates through the disk. 

I will discuss the morphology of these boundary layers and the evolution thereof following the initial sonic instability, before moving onto the details of wave-driven transport in Section \ref{sec:transport}. Figure \ref{fig:spacetime} first illustrates qualitatively how the angular velocity profiles of these disk-star systems evolve over time for a trio of $\mathcal{M}_*=8$ simulations with different thermodynamic treatments: in each case the outer regions of the initially non-rotating star spin up, gaining angular momentum from the inner edge of the accretion disk. In the isothermal and adiabatic cases, wave-driven angular momentum transport continues to appreciably alter the structure of the boundary layer throughout the first $\sim100$ orbital periods at the stellar surface, while in the $\beta=0.1$ disk this transport largely subsides by $\sim40$ orbital periods. 

Because these simulations employ neither viscosity nor magnetic fields, the large-scale structure of these disks tends to evolve very slowly over time after initial bursts of activity, for isothermal disks and even more so in the presence of cooling, although in most cases ($\beta\neq1$) waves remain appreciable until the end of each simulation.

The sequence of wave quasi-action ($rv_r\sqrt{\Sigma}$) snapshots in Figure \ref{fig:timesurvey} illustrates the persistence of shear-instigated acoustic waves throughout the disk until late times, although waves in the $\mathcal{M}_*=8,\,\beta=10$ visibly damp over time to a greater extent than the adiabatic and $\beta=10^{-2}$ disks, as expected based on the discussion in Section \ref{sec:diskdisp}. Also as presaged earlier, at later times the $\beta=10$ disk is occupied predominantly by lower-$m$ modes, as the higher-$m$ modes present at earlier times have gradually damped. This is in stark contrast to the adiabatic and $\beta=10^{-2}$ simulations portrayed in Figure \ref{fig:timesurvey}, which seem to support higher-$m$ modes until later times. 

Wave-driven transport in isothermal disks largely subsides after the first hundred or so orbital periods, although waves continue to propagate and transport angular momentum at a slower rate throughout each simulation; on the other hand, the amplitudes of these waves decrease dramatically in cooled disks, as illustrated in Figure \ref{fig:timesurvey}. However, this might not necessarily hold in cases where the angular momentum near the disk-star interface is replenished over time. 

To assess the systematicity of these trends and to investigate variation as a function of Mach number, Figure \ref{fig:machsurvey} shows how the waves in $\mathcal{M}_*=6$ and $\mathcal{M}_*=10$ disks evolve over time for each thermodynamic treatment. Figure \ref{fig:machsurvey} uses a color scale extending to much lower amplitudes in order to illustrate both that waves persist even in the $\beta=1$ disks (as these variations would be imperceptible using the color scale of Figure \ref{fig:timesurvey}), although the fading at $\beta=0.1$ and $\beta=10$ is still noticeable. Comparing the snapshots taken at $t = 100\pi\Omega_0^{-1}$ to those taken at $t = 400\pi\Omega_0^{-1}$, the transition of these systems from higher-wavenumber to lower-wavenumber configurations is generally apparent, especially for for cooling rates $\beta=1$ and $\beta=0.1$. Additionally, Figure \ref{fig:machsurvey} illustrates how cooler disks can support higher-wavenumber modes, even at very late times, thanks to modes with $m \lesssim \mathcal{M}_*\kappa/\Omega$ experiencing vanishingly small damping.\footnote{The values of $\kappa/\Omega$ typically range from $1$ to $2$ for Keplerian and rigidly-rotating disks respectively, which are the extremes expected to be relevant to these systems as illustrated by Figure \ref{fig:spacetime}.} Most disks support modes with $m>\mathcal{M}_*\kappa/\Omega$ at early times, but at later times only a subset of the isothermal and adiabatic disks support prominent higher-$m$ modes.

Figure \ref{fig:diskmodes} provides a complimentary view of the modes present in each disk, showing which modes (calculated using a Fourier decomposition of the radial velocity) are present over time. Although modes with a wide variety of wavenumbers are present at early times, those with $m\gtrsim \mathcal{M}_*\kappa/\Omega$ tend to die out in cooled disk by $t\sim200\pi\Omega^{-1}$; even in the $m\lesssim \mathcal{M}_*\kappa/\Omega$ regime, lower-$m$ modes are damped less strongly than those with higher $m$, and some cases such as those with $\beta=0.1$ gradually march towards lower $m$ over time. 
Some low-$m$ features develop at late times in isothermal and adiabatic disks as well. However, these are often due to the development of vortices in the boundary layer, which beget vortex-driven modes throughout the disk \citep{2021MNRAS.508.1842D,2022MNRAS.509..440C}.

Some cooled disks also develop vortex-driven modes. One visible example of a vortex-launched wave can be found in the $t = 400\pi\Omega_0^{-1}$ snapshot of the $\mathcal{M}_*=8$, $\beta=10$ simulation depicted in Figure \ref{fig:timesurvey}. Because these modes often have low-wavenumbers, they hold promise to operate even at late times. Additionally, in all cases with finite $\beta,$ cooling was always fast enough to prevent fluid heating from affecting the equilibrium disk structure. In the simulation of an adiabatic $\mathcal{M}_*=10$ disk, weak shocks during the initial stage of the instability slightly heated the star and slightly altering the background density profile, but by $t\sim 40\pi\Omega_0^{-1}$ the background density profile of the star became effectively constant. In each adiabatic case without cooling, the disk slowly heated up over the course of each simulation thanks to weak shocks.

\begin{figure}
\centering
\includegraphics[width=\linewidth]{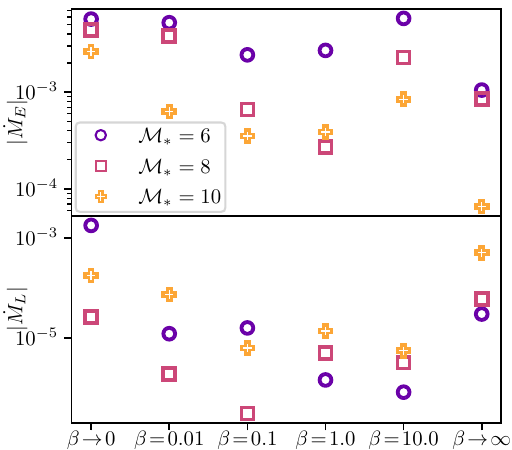}
\caption{Average accretion rates through the boundary layer, averaged over the range of radii $0.99\leq r/r_*\leq1.01$, since the steep increase in density within the star leads to fairly negligible mass flux through inner regions. The top panel plots the accretion rate at early times, averaged between $t=10\pi\Omega_0^{-1}$ and $t=100\pi\Omega_0^{-1}$, and the bottom panel plots the late-time accretion rates, averaged between $t=400\pi\Omega_0^{-1}$ and $t=600\pi\Omega_0^{-1}$. The star in the $\beta\rightarrow\infty,\,\mathcal{M}_*=10$ simulation was heated during the initial stages of the acoustic instability, and the increase in the pressure scale height somewhat counteracted accretion onto the star on average. At early times, cooled disks accrete only slightly less onto the star, but at late times the difference between cooled and uncooled disks is more stark.}
\label{fig:accrates}
\end{figure}

\begin{figure}
\centering
\includegraphics[width=\linewidth]{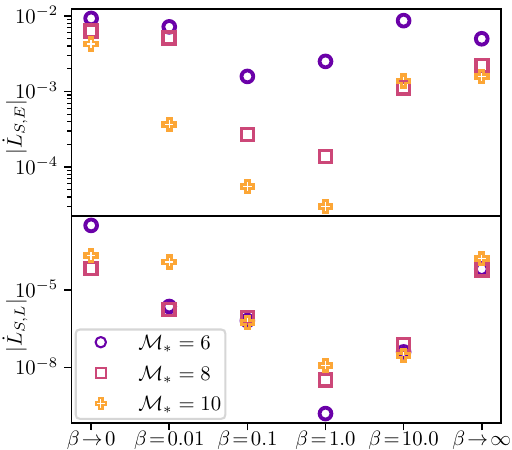}
\caption{Angular momentum currents (due to stresses) into the star, averaged over the range of radii $0.9\leq r/r_*\leq0.95$. The top panel plots the angular momentum current at early times, averaged between $t=10\pi\Omega_0^{-1}$ and $t=100\pi\Omega_0^{-1}$, and the bottom panel plots the late-time accretion rates, averaged between $t=400\pi\Omega_0^{-1}$ and $t=600\pi\Omega_0^{-1}$. Cooled disks exhibit weaker angular momentum transport at both early and late times, although this is in part because initial instabilities dissipate more quickly, as illustrated in Figure \ref{fig:spacetime}. At late times the disparity between cooled and uncooled disks is even more apparent than it was when viewing the accretion rate.}
\label{fig:currents}
\end{figure}

\subsection{Transport}\label{sec:transport}
With the effects of cooling and the passage of time on these disk-star systems in mind, we can turn to the transport of mass and angular momentum from accretion disks onto accreting stars. Because these simulations include neither viscosity nor magnetic fields, their phenomena are fundamentally transient, as illustrated in Figure \ref{fig:spacetime}. Transport of mass and angular momentum is most significant at early times, shortly after acoustic instabilities develop in the boundary layer, but continues until the end of each simulation, albeit in an extremely subdued state in the cases with $\beta$ closer to unity. Although accretion of mass and angular momentum are of particular astrophysical interest, whether the early- or late-time dynamics are more applicable to realistic systems remains to be seen.

The accretion rates through from the disk onto the star in each simulation are displayed in Figure \ref{fig:accrates}, averaged over both late and early times. The earlier averaging window, between the fifth and fiftieth orbital periods at the stellar surface, captures the strongest transport immediately after acoustic instabilities precipitate in the boundary layer. Because the temporal extent of this phase varied from simulation to simulation, the general trends in these time-averaged quantities are more useful than their precise values. As expected based on the preceding discussion, disks cooled with $\beta\sim1$ accrete onto the star at much slower rates, at both early and late times. Accretion rates are typically much lower at later times, by $\sim1-3$ orders of magnitude. It is worth keeping in mind that the accretion rate shown in Figure \ref{fig:accrates} includes mass transport by advection. The consequences of these terms are most apparent in the adiabatic $\mathcal{M}_*=10$ disk, which heats up considerably during the strongest stages of wave transport. That heating leads to a slight increase in the pressure scale height, causing the stellar atmosphere to puff outwards and counteracting the accretion of matter from the disk due to weakly-shocking waves. 

Many of the same trends appear in Figure \ref{fig:currents}, which displays the angular momentum current into the star due to waves, averaged over the same early and late time intervals as Figure \ref{fig:accrates}. However, cooling affects wave-driven angular momentum transport to a far greater extent than it did the accretion rate. In particular, in the $\beta=1$ disks, the rate of angular momentum transport through the boundary layer occurs at roughly three orders of magnitude lower a rate than in the isothermal and adiabatic disks at late times. This is a natural consequence of the damping of waves by cooling. 

\section{Discussion}\label{sec:discussion}
\subsection{Astrophysical Implications}
The characteristic cooling rate of an accretion disk around a given astrophysical object will strongly influence the evolution of that object over time. Disks with cooling timescales closer to the local dynamical timescale will accrete less efficiently onto a central object, and take longer to increase its angular momentum and rate of spin. This may help solve the fairly ubiquitous challenge of producing so few stars and planets with near-critical rotation rates. On one hand, if all of the angular momentum carried by the disk as it approaches the surface of an accreting object, many stars and planets that have grown via accretion would spin up to the point of shedding mass from their surfaces \citep[e.g.,][]{2008ApJ...685.1220M}; on the other hand, both Jupiter and Saturn both rotate at about a third of this critical rate, the limited set of measured exoplanets spins suggest rotation rates of $\sim0.08-0.3$ times the critical value \citep{2018NatAs...2..138B}, and the fastest-spinning millisecond pulsar also rotates at a subcritical rate \citep{2006Sci...311.1901H}.\footnote{Notably, if the accreting object is imbued with a strong magnetic field, magnetic braking against the disk can slow stellar spins. This process is thought to play a prominent role in setting the spins of T Tauri stars \citep[e.g.,][]{1996MNRAS.280..458A}. If, during formation, Jupiter possessed magnetic fields orders of magnitude stronger than it does today \citep{1974JGR....79.3501S}, then magnetic braking could also explain its current spin rate \citep{2018AJ....155..178B}.}

Previously, studies have investigated, through both simulations and linear theory, the waning efficacy of wave-driven transport through boundary layers as the accreting object rotates at higher and higher rates, which can be attributed to the development of Rossby modes within the star that tend to transport angular momentum outward \citep{2021MNRAS.508.1842D,2023ApJ...945..165F}. Disk thermodynamics may also play a crucial role, as the greatly diminished rate of angular momentum transfer between the disk and star at near-unity $\beta$ (Figure \ref{fig:currents}) would make spinning-up objects to near critical rates an extended and arduous process. Qualitatively, it appears that ratios of the cooling timescale to the dynamical timescale below $\beta \lesssim 10^{-2}$ or above $\beta \gtrsim 10^2$ are necessary to achieve degrees of angular momentum transport comparable to isothermal or uncooled adiabatic disks. 

However, this limited efficiency of transport may also exacerbate challenges in accretion. Although waves generated by acoustics instabilities in the boundary layers between disks and accreting objects are a promising transport mechanism, able to act in this region where the MRI ceases to operate, even isothermal disks may not be able to generate strong enough waves to keep pace with MRI operating in the outer disk, leading to a buildup of mass and angular momentum in a ring in the inner edge of the disk \citep{2018MNRAS.479.1528B}. Moderate cooling rates might worsen this situation, possibly leaving the problem of how accretion physically proceeds from disk to central object further from solution.

The analytic estimates in Section \ref{sec:diskdisp} and the simulations in Section \ref{sec:results} suggest that over time, cooling and thermal relaxation should damp higher-frequency modes, such that at late times the disk is populated by modes with $m\lesssim \mathcal{M} \kappa/\Omega$. The frequency of these waves is typically sub-Keplerian, very roughly with $m\Omega_p\sim\Omega-\kappa/m$. These low-frequency pressure waves can readily couple to gravity waves within the star \citep{2017ApJ...835..238B}, which propagate with frequencies below the Brunt-V{\"a}is{\"a}l{\"a} frequency and is $N\approx \mathcal{M}\sqrt{\gamma-1}\Omega_0$ near the stellar surface. Indeed, the mode measurements used to construct Figure \ref{fig:diskmodes} suggest that most of the significant modes within the stars and disks propagate at below this frequency. As discussed in \citet{2016ApJ...817...62P} and \citet{2017ApJ...835..238B}, these oscillations are good candidates for progenitors of quasi-periodic oscillations in low-mass X-ray binaries \citep[e.g,][]{2003A&A...410..217G,2005AN....326..812G} and dwarf nova oscillations in cataclysmic variables \citep[e.g.,][]{2004ApJ...616L.155P,2005ASPC..330..227W}.

\subsection{Caveats}\label{sec:caveatEmptor}
When interpreting the analytical estimates and numerical calculations presented in this work, one should not be ignorant to their nature, limitations, and the assumptions upon which they rely.

Although the parameterized cooling treatment of Equation (\ref{eq:cooling}) is useful for gaining general physical insight, it is also not strictly applicable to any astrophysical system. One approach taken in studies of protoplanetary disks, employing analytic in-plane and surface cooling functions \citep[e.g.,][]{2020ApJ...904..121M} or solving the equations of radiation transport in the flux-limited diffusion approximation \citep{1981ApJ...248..321L} in concert with physically motivated heating and cooling terms \citep[e.g.,][]{2024MNRAS.528.6130Z}, might provide an avenue towards more realistic studies of how stars and planets accrete. \citet{2015A&A...579A..54H} has confirmed that the sonic instability still operates under these conditions. Detailed studies of cataclysmic variables or X-ray bursts would require additional sources of heating from nuclear reactions, if not also more sophisticated treatments of radiative transfer. 

An additional limitation of the estimates presented in Section \ref{sec:diskdisp} is that the tendency for cooling to favor low-$m$ modes suggested by Equation (\ref{eq:simplecooldisk}) will eventually invalidate the high-$m$ assumption used to derive that dispersion relation. However, since this trend also came to pass in the simulations presented in Section \ref{sec:results}, this limitation may not be too severe. 

Another limitation of the simulations presented in Section \ref{sec:results} is their two-dimensional nature. On one hand, as demonstrated by \citet{2013ApJ...770...67B} through both analytic estimates and simulations, the acoustic instability and resulting wave-driven transport are essentially the same in two-dimension and three-dimensional disks, regardless of vertical stratification. On the other hand, in three dimensions the boundary layer could be susceptible to the Kelvin-Helmholtz instability, because even though this instability is suppressed at high Mach numbers \citep{1958JFM.....4..538M}, in three dimensions there will always exist a wavevector with respect to the flow velocity such that the projected Mach number is subcritical and the flow is unstable \citep{1963JFM....15..335F}. Although the conclusions regarding wave-driven transport in this work are probably robust, other mechanisms important to the development of boundary layers may have been omitted. Additionally, accounting for vertical stratification is necessary to study the spreading of accreted material across the stellar surface \citep[e.g.,][]{1999AstL...25..269I,2004ApJ...616L.155P,2016ApJ...817...62P}. 

Of potentially the most importance out of the limitations of this work, I have neglected the effects of viscosity and magnetic fields. Although these do not strongly affect the mechanics of wave-driven transport \citep[e.g.,][]{2013ApJ...770...68B,2015A&A...579A..54H,2018MNRAS.479.1528B}, they are necessary to continue bringing mass and angular momentum to the boundary layer over time. The dynamics studied in Section \ref{sec:results} were intrinsically transient. For example, at early times in some disks, mass and angular momentum transport  was only slightly weaker in cooled disks, and if this transient state happens to be more representative of the longer-timescale evolution of these systems, the effects of cooling may not be as dramatic. However, this remains to be seen.

\section{Summary}\label{sec:summary}
This work has investigated how differing thermodynamic effects, primarily the cooling rate, affect
boundary layer-mediated accretion onto stars and planets. This region where the angular frequency of gas increases rather than decreases is stable to the MRI, and instead shear-driven acoustic instabilities are a leading physical mechanism to transport material and angular momentum from the disk to the accreting object. The waves generated by this sonic instability are strongly affected by cooling. 

In the Newtonian cooling approximation, Section \ref{sec:methods} investigated the effects of cooling on wave propagation in general and within disks. Qualitatively, when the cooling timescale is comparable to the wave period, cooling very strongly damps waves. When cooling occurs comparatively quickly, the waves propagate nearly isothermally, and when the cooling timescale is longer many wave periods occur before cooling can appreciably affect their dynamics. In disks, this translates to very weak damping for waves with azimuthal wavenumbers $m\lesssim \mathcal{M}\kappa/\Omega$; these lower-frequency modes appear more likely to excite incompressible modes within the central object, which are good candidates for low-frequency variability such as dwarf nova oscillations. This analytical estimate was confirmed by the simulations in Section \ref{sec:results}, which also demonstrated the deleterious effects of cooling on mass and angular momentum transport --- most significantly at intermediate cooling timescales, but also over the entire range considered $(0.01\leq \Omega t_c \leq 10)$. 

\section*{Software}

\texttt{matplotlib} \citep{4160265}, \texttt{cmocean} \citep{cmocean}, \texttt{numpy} \citep{5725236},  \texttt{Disco} \citep{2016ApJS..226....2D}

\section*{Acknowledgments}
I am grateful for numerous discussions with Sasha Philippov, which helped to motivate and guide this work. I am also grateful for invigorating scientific discussions with Geoff Ryan, and his contributions to \texttt{Disco}. I also thank Tad Komacek and Sasha Philippov for providing feedback on a draft of this manuscript. I acknowledge support from NASA ADAP grants 80NSSC21K0649 and 80NSSC20K0288. The simulations presented in this work were conducted on the Zaratan cluster at the University of Maryland (http://hpcc.umd.edu).

\bibliographystyle{aasjournal}
\bibliography{references}
\end{document}